# Dielectric Barrier Discharge Actuators: Experimental and Numerical Study of Momentum Injection into Co-flow and Counter-flow Freestream

Anthony Tang[1], Nathan Li[1], Benjamin Price[1], Alexander Mamishev[2], Alberto Aliseda[1], Igor Novosselov[1,3,*]

[1]*Department of Mechanical Engineering, University of Washington, Seattle, U.S.A. 98195*

[2]*Department of Electrical and Computer Engineering, University of Washington, Seattle, U.S.A. 98195*

[3]*Institute for Nano-Engineered Systems, University of Washington, Seattle, U.S.A. 98195*

## ABSTRACT

Dielectric barrier discharge (DBD) plasma actuators can generate a wall jet without moving parts by interacting with ionized and neutral molecules in an electric field. The coupling between electrohydrodynamic (EHD), turbulence, inertial and viscous effects in the flow boundary layer remains poorly understood and requires investigation. We present an experimental investigation of momentum injection by DBD actuators into the free stream flow with Re = 35,000 and 75,000 in co-flow and counter-flow scenarios over a range of $V_{AC}$ = 12 kV - 19.5 kV peak-to-peak at a frequency of 2 kHz. In the co-flow configuration, the DBD actuator injects momentum into the boundary layer. In co-flow, the momentum injection results in the thinning boundary layer, while in the counter-flow configuration, flow separation can occur. For the tested condition, a separation bubble is observed at Re = 35,000. The momentum displacement in the counter-flow configuration is six times greater than the EHD jet momentum in a quiescent environment. Both co-flow and counter-flow momentum injections show diminishing effects with increasing external velocities. This work highlights that the resulting flow pattern is not a simple superposition of the EHD jet and the free stream but is determined by the coupling of inertial, viscous, and Coulombic effects in the EHD-driven wall jet and the external flow. The velocity profiles and momentum measurements presented here can be used to validate numerical models and inform the design of DBD actuators for active flow control.

## 1. INTRODUCTION

Non-thermal plasma devices have been proposed as actuators for active flow control [1-7]; these plasma actuators have the potential to instantaneously change flow profiles while staying silent and compact [8-10]. Corona discharge or dielectric barrier discharge (DBD), actuators generate ions when the electric field exceeds the dielectric strength of the working fluid. The interaction between free ions, accelerated by the electric field, the working fluid, and walls can be utilized in aerodynamic drag reduction [11-13], lift augmentation [10, 14], separation control [15, 16], and electric propulsion [17-20]. Despite lower electromechanical efficiency than corona-driven actuators, DBD actuators are more stable and can effectively provide a consistent electrohydrodynamic (EHD) forcing [4, 9]. Current DBD applications are limited to flow control at low-speed conditions due to their relatively weak EHD forces [17, 21, 22]. Scientific studies have explored these multiphysics phenomena to optimize electrical and mechanical effects in DBD systems [16, 23-29]. Modeling DBD discharge from first principles is cost-prohibitive due to the large separation of timescales, so studies of corona-driven ionic flows can be relevant to gain insight into the flow-ion interactions [7, 25, 27, 30]. Early numerical efforts led to the development of simplified DBD ion injection models, including the Suzen

---

* ivn@uw.edu

& Huang [31], Orlov [32], and Shyy [33] models that were able to predict quiescent EHD flow through the interactions of positive and negative ion charges and the external environment [9, 29]. More recent work has pushed past these earlier models to numerically explore the performance of DBDs, such as the evolution of the velocity field in the DBD-driven jet [28]. In addition, other strategies have been proposed for more robust, computationally-lean momentum injection models for EHD-driven jets [34]. Nearly all these models were developed and validated in quiescent conditions and are yet to be tested in an external flow condition with significant inertial and viscous flow interactions. Several authors have noted and presented varying success in quiescent conditions with different numerical approaches, including direct numerical simulations and turbulence RANS closure models [35-37].

Most reports describing EHD-flow interaction are currently limited to analysis of electroconvective instabilities at very low Reynolds numbers. Electroconvection (EC) phenomenon was first reported by G. I. Taylor in 1966, describing cellular convection in the liquid droplet [38]. Since then, EC has been observed in other systems where electric force interacts with fluids. In nonequilibrium EHD systems [38-59], poorly conductive leaky dielectric fluid acquires unipolar charge injection at the surface interface in response to the electric field. In charge-neutral electrokinetic (EK) systems, EC is triggered by the electro-osmotic slip of electrolyte in the electric double layer at membrane surfaces [60-71]. In 3D shear flow, the EHD addition to crossflow forms streamwise rolling patterns as observed numerically [57, 72-74] and experimentally [75, 76]. 2D and 3D flow analysis of the DBD momentum injection in the shear flow is missing from the published literature. A mechanistic understanding of the interaction between discharge and fluid flow in the presence of an external flow is needed to inform the development of DBD actuators for active flow control.

To maximize the effect of DBD actuators, recent experimental work varied actuator geometries such as electrode shapes, number of electrodes, and their placement on an aerodynamic surface. However, most of the fundamental EHD studies were performed in a quiescent environment; these actuators have not been well studied in the presence of an external freestream flow, especially at low to moderate velocity relevant to flow separation control [77, 78]. Several airfoil and small-scale unmanned aerial vehicle (UAV) studies have explored the ability of DBD actuators to manipulate lift and drag forces; however, these studies did not provide insight into the fluid flowfield and the underlying physics responsible for the lift and drag changes [79, 80]. The two traditional external flow conditions include co-flow, when the jet direction is the same as the external flow, and counter-flow when the momentum injection is opposite to the external flow. Pereira et al. reported force measurements in both co- and counter-flow DBD injection. They found that the total EHD thrust (or the difference between EHD thrust and shear stress at the surface) was identical for both co-flow and counter-flow [81]. However, the range of freestream velocities (10 – 60 m/s) in increments of 10 m/s did not address the regime where the EHD velocity equals the external flow (0 – 10 m/s) [81]. Probing the underlying fluid dynamics requires measurement of the velocity profiles near the momentum injection location. Bernard et al. reported on the velocity profiles of DBD actuators in co-flow and found that the effect of the DBD injection diminished at higher external velocities; however, the authors did not investigate counter-flow conditions [82]. The literature does not report experimental work characterizing velocity profiles in the counter-flow momentum injection by DBDs.

Over the past decade, several studies have been conducted on DBD mounted to various airfoils [14, 83, 84], multi-element airfoils [85], flaps [86, 87], and full-scale or near full-scale aircraft [17, 21]. In all studies, the DBD actuator demonstrated its ability to change aerodynamic performance by increasing airfoil lift, decreasing drag, or changing pitching moment. Gu et al. employed a simplified ion injection model to explore the effects of DBD actuators on Gurney flaps by modeling actuators on the pressure and suction side of the trailing edge of an airfoil [36]. Multiple DBD array systems have been tested with moderate success; many of these studies found the need to balance the simultaneous interactions between jets acting in opposite directions [77, 88]. Counter-flow momentum injection potentially manipulates the flow more efficiently by increasing drag, decreasing lift, and changing the pitching moment on an aerodynamic surface.

This study explores the performance and effect of EHD momentum injection by the DBD actuator at $U_\infty = 5$ m/s and $U_\infty = 11$ m/s in co-flow and counter-flow. The AC voltage was varied in the range $V_{AC} = 12$ kV – 19.5 kV, and the frequency was constant at 2 kHz. This study is the first to explore the fluid characteristics of DBD-driven flow in counter-flow and quantify the onset separation due to an adverse pressure gradient. Finally, a simple momentum injection model based on the empirically derived DBD discharge in a quiescent environment was tested in external flow computational fluid dynamics (CFD) simulation. This work provides insight into the interaction between the DBD flow and a freestream flow over a flat plate, informing the potential placement of actuators on airfoils and providing validation for numerical studies.

## 2. EXPERIMENTAL SETUP AND DIAGNOSTICS

Traditional metrics to characterize plasma actuators' performance include current, power consumption, forces on the surfaces, and DBD jet velocity. A current measurement shows a superposition of capacitive current, discharge current, and noise. The capacitive current is filtered out or ignored because it corresponds to the transiently stored energy in the dielectric or air, not the energy used to accelerate the fluid [5]. The discharge current indicates the amount of charged species that can participate in the energy transfer to fluid motion. The discharge current comprises of numerous peaks in the positive-going cycle due to streamer propagation with the addition of glow discharge during the negative-going cycle [89]. Recent research characterized the relationship between DBD discharge, capacitance, power consumption, and DBD actuator performance [90, 91]. High-resolution measurements have shown that both positive and negative half-cycles contribute to EHD force, and their relevant contributions are the topic of active scientific discussions [5, 92, 93]. Velocity measurements can be obtained by pitot tubes or particle imaging velocimetry (PIV); these measurements characterize momentum transferred from charged species to neutral molecules. While PIV measurements can capture an entire fluid field, integrating a PIV system with DBD-induced flow can be challenging. In addition, there is a risk of tracer particles being charged and interacting with the electric field, i.e., not following the flow streamlines. DBD wall jet similarity analysis was recently proposed [94]; however, additional experimental data is needed to perform a robust analysis.

### 2.1. Wind Tunnel

Our measurements were conducted in an open-circuit wind tunnel with a 100mm x 100mm cross-sectional test section. The wind tunnel consists of an inlet section followed by a 1000 mm long section and a test section allowing for testing the DBD actuator. The DBD actuator plate surface rests parallel to the bottom wall of the wind tunnel, see Figure 1. The inlet section comprises four 120 mm x 120 mm fans with inlet cowls, a large honeycomb screen, and a contraction cone. The contraction cone with a 9:1 contraction ratio results in a 100mm x 100mm wind tunnel section constructed of plexiglass. An aluminum extrusion frame supports the wind tunnel section. As described below, the velocity measurements were obtained using a custom glass pitot tube with a 0.4mm ID and 0.5mm OD. The boundary layer height ($\delta_{99}$) at $U_\infty = 5$ m/s external flow was measured at the location of the actuator to be ~8.0 mm. Using the height of the wind tunnel and the characteristic length; the Re is 35,000 at $U_\infty = 5$ m/s and 75,000 at $U_\infty = 11$ m/s. The turbulence intensity is ~1% measured using a calibrated hot-wire anemometer. The hot-wire anemometer was not used for velocity measurements of the DBD actuator as the hot-wire wires and electronics would introduce a risk of arc discharge from the high-voltage electrode.

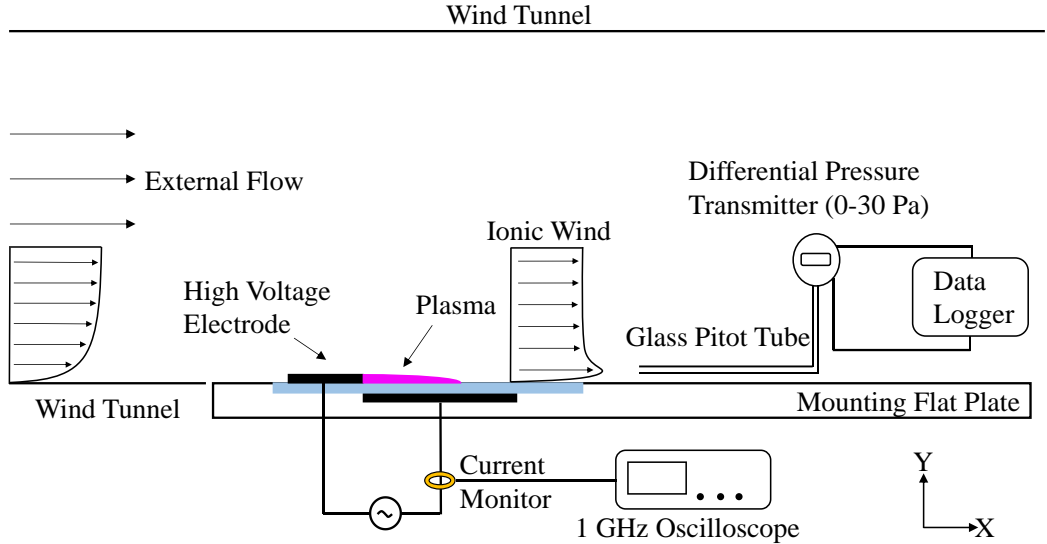

**Figure 1. Schematic of the experimental setup, DBD actuator is mounted on an acrylic glass plate flushed with the wind tunnel bottom. The blue region is the dielectric layer separating the electrodes.**

### 2.2. DBD actuator

The DBD actuator is comprised of two electrodes separated by a thin dielectric barrier, as shown in Figure 1, similar to previously published work [95]. When high voltage is applied to the active electrode, the electric field is strongest at the edge of the active electrode, resulting in plasma generation [92, 96, 97]. The electrodes' thickness and the dielectric media can impact the actuator's performance [97-101]. A straight-edged DBD actuator with a spanwise uniform electric field produces a two-dimensional forcing on the fluid, resulting in a planar jet. Other actuator designs have been considered, including serrated electrodes that produce a three-dimensional force onto the flow field [102-104]. The spanwise length or width of the electrodes serves as a nominal reference length in the analysis [5]. The DBD actuator was installed on the 6″ by 8″ acrylic plate for this study. The dielectric material used in this study is Kapton (~3.5 dielectric constant at 1 kHz). Each actuator has one layer of ~0.075mm Kapton-FN (FEP layered Kapton) and four layers of 1 mil Kapton-HN with a total thickness of ~ 0.4mm (including the adhesive and FEP layers). The ground electrode (copper, 50 µm thick, 25 mm long, 110 mm wide) is flush-mounted on the acrylic plate. The upper electrode (copper, 50 µm thick, 15 mm long, 110 mm wide) is glued onto the top of the Kapton dielectric layer. Both electrodes have straight edges producing a uniform spanwise discharge. The active and ground electrodes' edges are aligned with each other, i.e., there is no overlap or gap between the electrodes in the x-direction. The air-exposed HV electrode is connected to a Trek 615-10 (PM04014) power supply that provides up to 20 kV (peak-to-peak) AC voltage.

### 2.3. Electrical Measurements

The electric current in the DBD circuit is a superposition of a capacitive current and a discharge current. The discharge current is associated with plasma microdischarges, and they appear as a series of fast current pulses [105], as shown in Figure 2(a). DBD current is measured using a 200 MHz bandwidth non-intrusive Pearson 2877 current monitor with a rise time of 2 ns. The current monitor is connected to a Tektronix DPO 7054 oscilloscope with a sampling rate of 1 GS/s to resolve 500 MHz. These conditions are essential for accurately capturing individual discharges that have been shown to occur over a 30 ns duration on average [106]. The high bandwidth and the sampling rate minimize the noise during the current measurements and can be used to compute the time-averaged electrical power [52]. To determine the currents associated with the plasma micro-discharges, determining the capacitive current through analytical methods has been explored [32, 107]; removing the capacitive current through signal processing methods, including low-pass filters or Fast-Fourier Transform (FFT) has also been attempted [93, 105, 106].

To determine the power consumed by the actuator, a charge-voltage Lissajous curve is created by introducing a capacitor between the grounded electrode and the ground [108, 109]. The Integrating charge-voltage relationship multiplied by the frequency yields the total power usage of the DBD system. The time-averaged electrical power consumed by the actuator is computed as

$$W_{elec} = f_{AC} \int_{t^*=0}^{t^*=1} V dQ, \qquad (1)$$

where $f_{AC}$ is the frequency of the applied voltage in Hz, and $V$ and $Q$ are the voltage and charge at each point in the period. The normalized time ($t^*$) represents a single cycle. We compute the averaged resulting power from at least four separate periods to reduce the noise impact. In the wind tunnel study by Pereira et al., the DBD actuator in co-flow and counter-flow was found not to have significantly different electrical characteristics [81]. Figure 2(a) below shows a typical DBD current measurement with a voltage curve. Figure 2(b) shows the representative filtered Lissajous curve of four consecutive discharge cycles. These data were used to determine the power of the DBD actuator as a function of the operating condition.

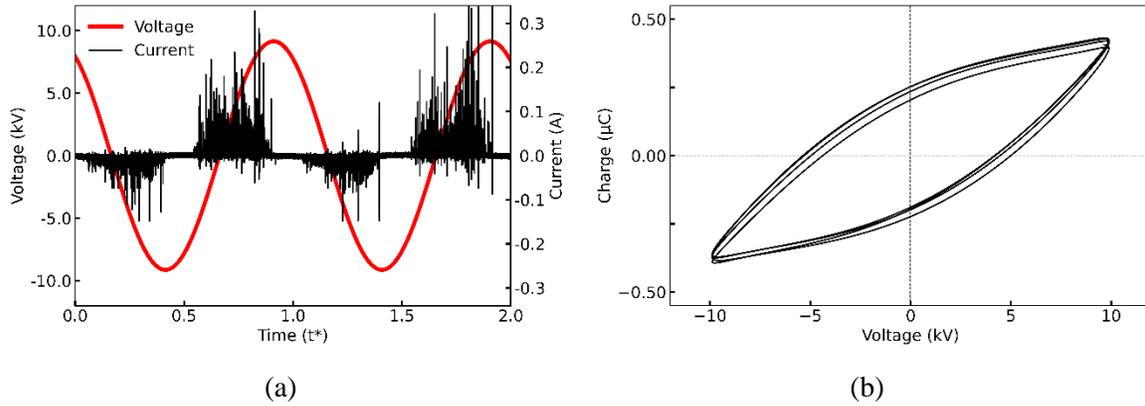

(a) (b)
**Figure 2. (a) Typical DBD current with voltage signals at 18 kV (p-p) and 2 kHz applied frequency (b) 4 consecutive Q-V discharge cycles measured from a 100 nF capacitor.**

### 2.4. Wall jet and momentum displacement characterization

The flowfield induced by EHD depends on the plasma actuators' configuration and operational conditions. We employed a custom-made glass pitot tube with a 0.4 mm inner diameter and 0.5 mm outside diameter to measure the time-averaged x - velocity profile. Compared to traditional stainless steel pitot tubes, the glass tube reduces electrical interaction with the discharge. This method has been previously used to characterize plasma actuators' performance [5, 95, 105, 110]. The pitot tube is mounted on an optical table and controlled on the x and y - axis by linear stages connected to an Ashcroft CXLdp differential pressure transmitter (0 – 25 Pa with 0.25% accuracy). The pressure transducer outputs a 4 – 20 mA current, linear in its pressure range, and it was placed in series with a 1.5 kΩ resistor. The pressure within the pitot tube equilibrated nearly instantly after changing the flow condition. The voltage across the resistor is recorded for at least 30 seconds with a Hydra Data Logger II. With the time-averaged pressure ($P$), a time-averaged wind velocity ($v$) is calculated using

Bernoulli's equation with a calibration correction factor ($C$) that is characteristic for a custom pitot tube expressed as

$$\Delta P = C\rho v^2, \quad (2)$$

where $\rho$ is the fluid density. In our experiments, the typical velocity measurements had a standard deviation $< 0.02$ m s$^{-1}$ over a 30 s sampling period. X-velocity measurements are taken at varying x and y positions downstream and upstream on the active electrode edge. At each x-position, the y-velocity profile is obtained from the surface to 10 mm above the plate at increments of 0.25 mm or 0.5 mm (at a higher location). The streamwise measurements were taken by holding a constant y-position and spanning in the x - direction at 0.5 mm intervals to complete the datasets over a regular measurement grid. Due to the pitot tube dimension, we could not capture velocity at y $<$ 0.25 mm. We assume the velocity is linear between the no-slip condition at y $=$ 0 mm and the data at y $=$ 0.25 mm for plotting purposes.

Considering a 2D control volume (with a spanwise unit length), integration of the streamwise velocity along a vertical profile in the wind tunnel with the DBD actuator provides the total mass flow rate per meter spanwise, $Q$, of the total system:

$$Q_{system} = \rho \int_{y=0}^{y=h} U(y)dy, \quad (3)$$

where $U(y)$ is the velocity measured along the vertical profile at a constant x - location. Similarly, the system's total momentum can be found by integrating the square of the velocity along a vertical profile:

$$M_{system} = \rho \int_{y=0}^{y=h} U^2(y)dy. \quad (4)$$

To identify the momentum produced by only the DBD actuator, the same measurements were taken with the DBD actuator ON and OFF. The difference in momentum integrated from the wall to the point where the two velocity profiles intersect is the momentum injected by the DBD. Above the intersection of the two profiles, there is entrainment due to mass conservation, and this entrainment is confirmed by taking the velocity profile of the entire wind tunnel with and without actuation. The resulting momentum is expressed as

$$M_{DBD} = \rho \int_{y=0}^{y=h} U_{DBD\ ON}^2(y) - U_{DBD\ OFF}^2(y)dy \quad (5)$$

and this approach similarly holds for mass flow rate and mechanical power. The mechanical power of the system ($W_{mech}$) can be computed by

$$W_{mech} = \frac{1}{2}\rho L \int_{y=0}^{y=\infty} U^3(y)dy. \quad (6)$$

The derived values of mass flow rate, momentum, and power of the DBD in external flow are compared to the results of similar measurements in quiescent flow.

### 3. RESULTS AND DISCUSSION

#### 3.1. Power Consumption

Figure 3 illustrates the power usage of the DBD for a range of operating voltages, external flow speed, and DBD actuator orientation. Integration of Lissajous Q-V curves yielded an average power consumption of the actuator. The power usage is normalized to the spanwise length of the actuator (0.11 m). In general, the power consumption increases quadratically with applied voltage, which is consistent with previous reports for AC DBD [5, 92, 95, 111] and the EHD flows driven by corona discharge [26, 30, 112]. Power usage between the AC cycles for any configuration had a maximum variance of ~ 8%, see Figure 3. These data were taken for the DBD actuator in quiescent, counter-flow, and co-flow conditions at the two external flow speeds. Pereira et al. also found power usage to vary less than 10% between co-flow and counter-flow forcing [81]. The magnitude of normalized power usage measured for this study is slightly higher than Pereira et al., which used a thicker 3mm dielectric; however, other studies, such as Forte et al., have found similar power usage for thinner thicknesses and noted that the increase in dielectric thickness decreases DBD capacitance, thus decreasing power usage [5, 92, 95].

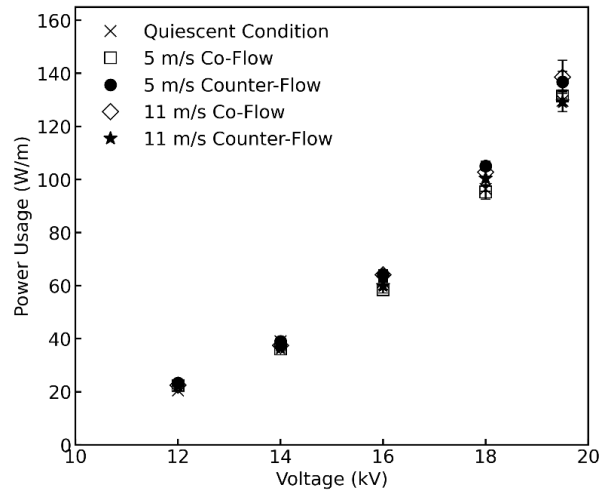

**Figure 3. DBD power usage at $U_\infty$ = 5 and $U_\infty$ = 11 m/s external flow in co-flow and counter-flow**

### 3.2. Operation in Quiescent Condition

First, the momentum injection of the AC DBD actuator is considered in a quiescent environment. Figure 4 shows the velocity profile of the DBD jet at x = 10 mm and x = 25 mm downstream of the active electrode edge [95]. Although the velocity decay and the spreading of the jet are apparent, as expected in wall jets, the presence of Coulombic forces associated with the DBD and the fact that the fluid is being accelerated over the fluid volume rather than the point source make the parameterization of the flow challenging. The addition of co- and counter-flow further complicates the analysis.

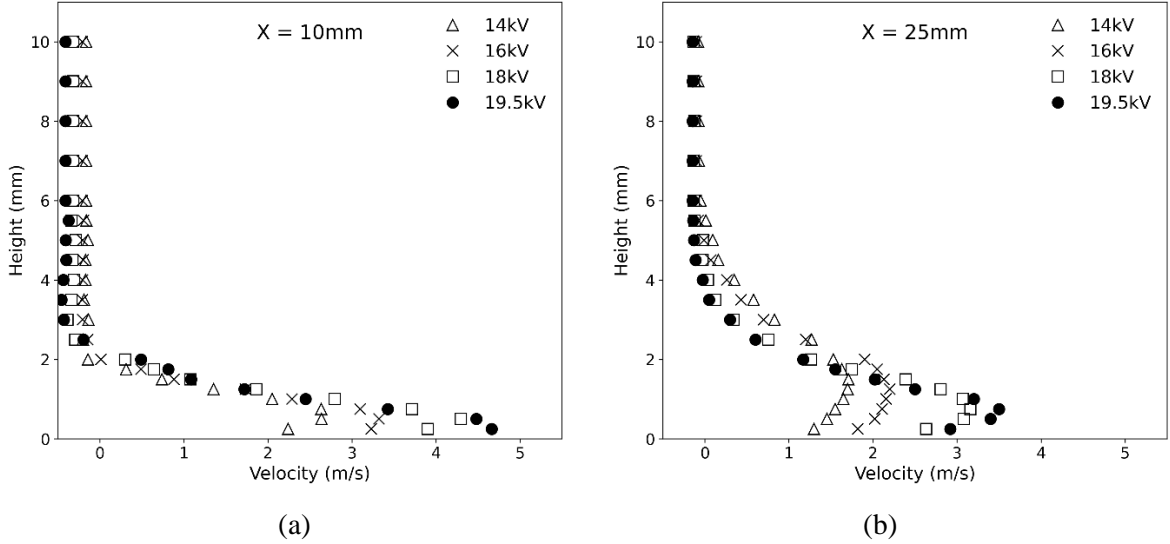

(a)                  (b)

**Figure** 4. **DBD with no external flow at x = 10 mm (a) and x = 25 mm (b) downstream. The maximum velocity at 19.5kV is ~ 4.7 m/s at y = 0.5 mm at x = 10 mm downstream.**

### 3.3. Co-flow EHD Momentum Injection

This section discusses the DBD actuator performance in co-flow over the range $V_{AC}$ = 14 – 19.5 kV, a frequency of 2 kHz, $U_\infty$ = 5 m/s (Re=35,000) and $U_\infty$ = 11 m/s (Re-75,000). The virtual origin and coordinate system are defined in Figure 5. The velocity profiles of the EHD jet are measured at x = 10 mm and x = 25 mm downstream of the active electrode edge.

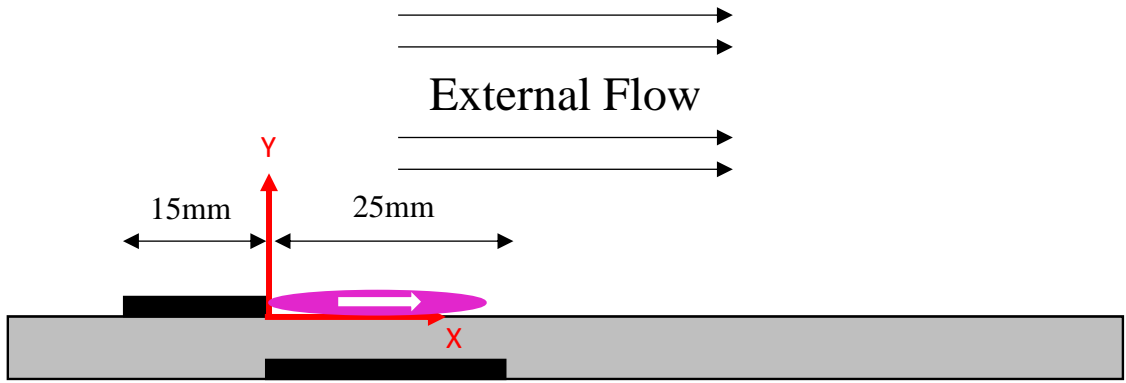

**Figure 5. DBD actuator in co-flow configuration, the first measurement is taken at x=10 mm to avoid plasma region disruption with pitot probe. The plasma region is colored purple.**

Due to EHD momentum injection, the velocity in the boundary layer increases; however, the difference in momentum displacement is affected by the thickness of the freestream boundary layer. Figure 6 shows the velocity data versus DBD voltage at two downstream locations. The dotted line is the wind tunnel velocity profile without DBD actuation. Note that in quiescent conditions (Figure 4), the DBD wall jet has a maximum velocity, $U_{max}$ ~ 4.8 m/s at x = 10 mm, y = 0.25 mm when $V_{AC}$ = 19.5 kV and $f_{AC}$ = 2 kHz, and the peak velocity of the jet decays quickly away, downstream and wall-normal, from this maximum. At $U_\infty$ = 5 m/s, EHD-induced velocities are comparable to the freestream, and the effects of DBD actuation are dominant. The increase in boundary layer velocity of ~ 2 m/s is nearly identical to that of Bernard et al. [82] for similar conditions. The $U_{max}$ ~ 5.4 m/s is greater than in quiescent condition (~ 4.8 m/s) but located at y = 0.75 mm for the same x-position, as viscous effects shift the location of $U_{max}$ away from the wall. At y = 0.25 mm, the U velocity is 5.3 m/s and is greater than that of the quiescent jet at y= 0.25 mm; thus, the viscous effects between the DBD jet and freestream can be seen as mixing and entraining fluid from the external freestream into the boundary layer. In co-flow, the DBD-induced momentum does not diffuse into the outer flow as quickly as in the quiescent environment, and this can be observed from the velocity profile at x = 25 mm. The interaction of the wall jet with the freestream means that its momentum does not diffuse into a quiescent environment but rather continues to mix and entrain fluid into the boundary layer, high-speed fluid from the freestream. This entrainment can be seen by a slight decrease in the external flow with the energized DBD actuator starting at approximately y = 3.75 mm. Conservation of mass in the wind tunnel means that as the boundary layer accelerates by the EHD-added momentum, the freestream velocity decreases slightly (the momentum thickness of the boundary layer is smaller). The complete velocity profile confirms that mass is conserved as the entrainment section eventually merges with the base wind tunnel profile.

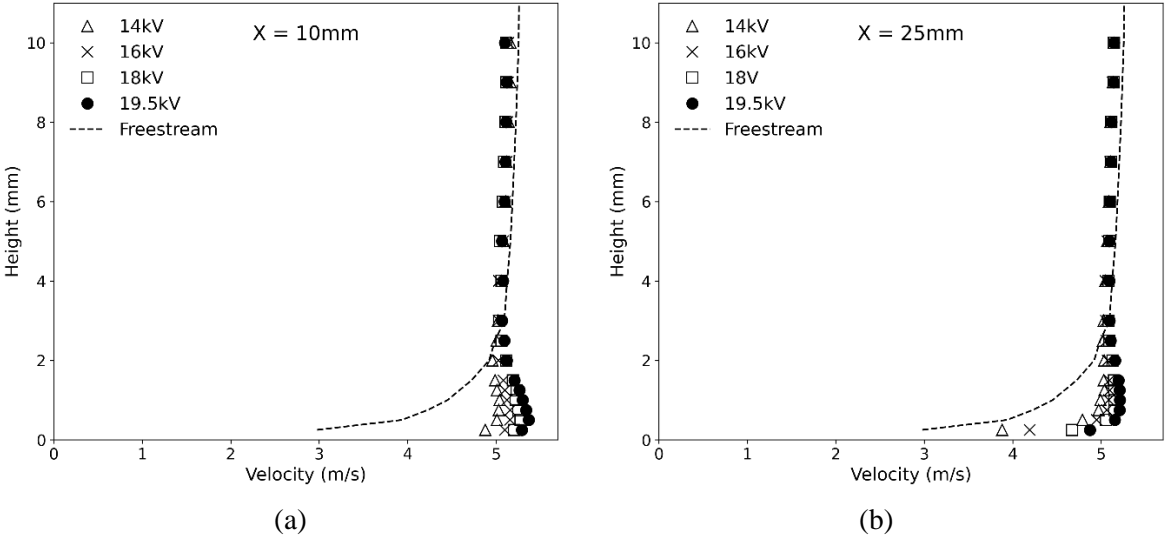

**Figure 6. DBD actuator in $U_\infty$ = 5 m/s co-flow at (a) x = 10 mm and (b) x = 25 mm downstream. The dashed line shows the freestream profile without plasma injection. The DBD voltage is varied in the 14kV-19.5kV range; the AC frequency is set constant at 2kHz.**

Figure 7 shows the effect of the EHD wall jet in the boundary layer at $U_\infty$ = 11 m/s. In this case, the EHD velocity is about half the freestream for the highest DBD settings. The effect of momentum injection is reduced, as the enhanced mixing in this higher Reynolds number case is more effective at spreading the effect of the EHD momentum injection throughout a thinner boundary layer. Even at maximum DBD power, the velocity increase is less than 1 m/s at x = 10 mm. At x = 25 mm, the effect of the EHD momentum injection is almost negligible. These results agree with Bernard et al. [82] for $U_\infty$ = 10 m/s. At higher external flows, the EHD momentum addition results in a lower overall impact on the boundary layer, as the enhanced mixing in the thin boundary layer rapidly restores the boundary layer profile to the un-actuated shape.

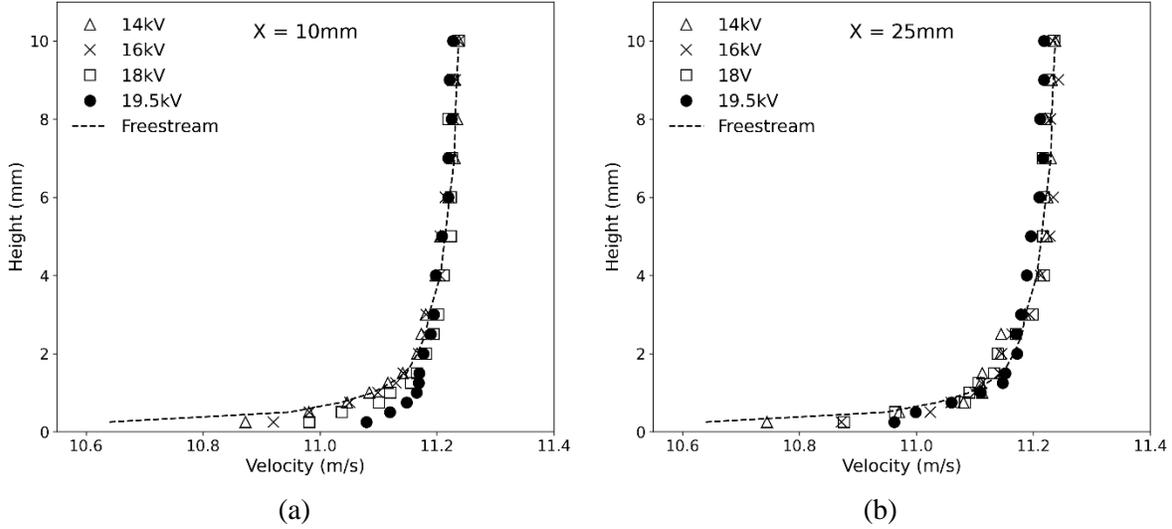

**Figure 7. DBD actuator in U∞ = 11 m/s co-flow at x = 10 mm (a) and x = 25 mm (b) downstream. The DBD voltage is varied in the 14kV-19.5kV range, the AC frequency is set constant at 2kHz.**

The EHD momentum addition cannot be treated as a linear superposition of the EHD jet in a quiescent environment and the momentum associated with the boundary layer of the free stream. For external flows compatible with EHD wall jet velocities, the momentum injection into the co-flow leads to effective boundary layer thinning; this effect diminishes at higher freestream velocities (higher Reynolds numbers and thinner boundary layers). The wall jet mixing is influenced by (i) interaction with the freestream and (ii) viscous wall shear increase in the viscous sublayer. This point is explored further in Figure 13.

### 3.4. Counter–Flow EHD jet

This section describes the behavior of counter-flow EHD jet at DBD $V_{AC}$ = 14 kV – 19.5 kV at $f_{AC}$ = 2 kHz and wind speeds of $U_\infty$ = 5 m/s and $U_\infty$ = 11 m/s. The virtual origin and coordinate system of the DBD in counter-flow are defined above in Figure 8. The datum for analysis is set at the plasma generation edge of the active electrode; however, the EHD momentum injection is now in the negative x-direction.

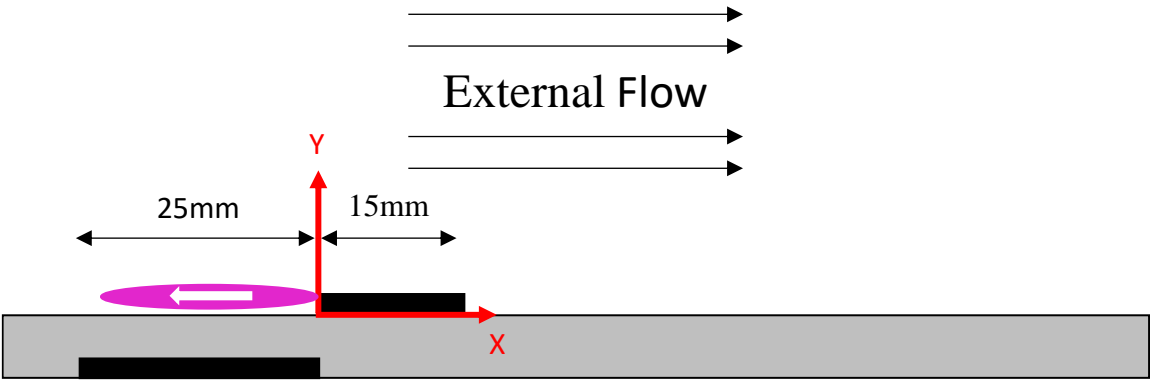

**Figure 8. DBD actuator in counter-flow configuration. The plasma region is colored purple.**

Figure 9 and Figure 10 show the velocity profiles for the EHD momentum injection into the counter-flow. The dotted line is the measured wind tunnel velocity profile without DBD actuation. At $U_\infty$ = 5 m/s, the velocity of the EHD jet has a similar magnitude to the external flow resulting in a significant adverse pressure gradient and the formation of a recirculation zone. The exact boundaries of the separation region are difficult to determine experimentally in the plasma region (x < 0 mm, y < 2mm) as the insertion of the pitot probe into the plasma interferes with the experiment, see Figure 8. However, to preserve continuity, the EHD jet must entrain fluid from above and behind the jet; thus, transects downstream and along x for constant y can be used to determine the boundaries of the separation bubble.

First, we examine the y-scan at the fixed x-position. Figure 9 and Figure 10 show the profiles at x = 10 mm (above the active electrode) and x = 25 mm for $U_\infty$ = 5 m/s and $U_\infty$ = 11 m/s, respectively. As with the co-flow experiments, the EHD jet strength is varied by varying $V_{AC}$ = 14 kV - 19.5 kV, while the AC frequency is 2kHz. For all voltages, the DBD in counter-flow creates a more significant deficit in the boundary layer than its co-flow counterpart, e.g., the counter-flow EHD jet creates a $\Delta U > 5$ m/s at $V_{AC}$ = 19.5 kV compared to the $\Delta U \sim 2$ m/s in the co-flow case. Figure 9 (a) also shows that in counter-flow cases at 18kV and 19.5kV, the wall shear stress changes direction due to the increased strength of the EHD jet. In the co-flow scenario and lower voltage counter-flow configurations, the wall shear stress remains opposite of the freestream direction. Note that the maximum negative velocity is likely located in the EHD momentum injection region (x= - 10 mm – 0 mm). However, measurements could not be taken within the plasma region due to the plasma interactions with the pitot tube. Figure 9 shows that in the $V_{AC}$ = 19.5 kV case, the separation bubble extends past x = 10mm downstream of the active electrode edge, while other conditions exhibit flow reattachment. The flow is fully attached at x = 25 mm; however, the pressure gradient in the flow boundary layer has not yet recovered.

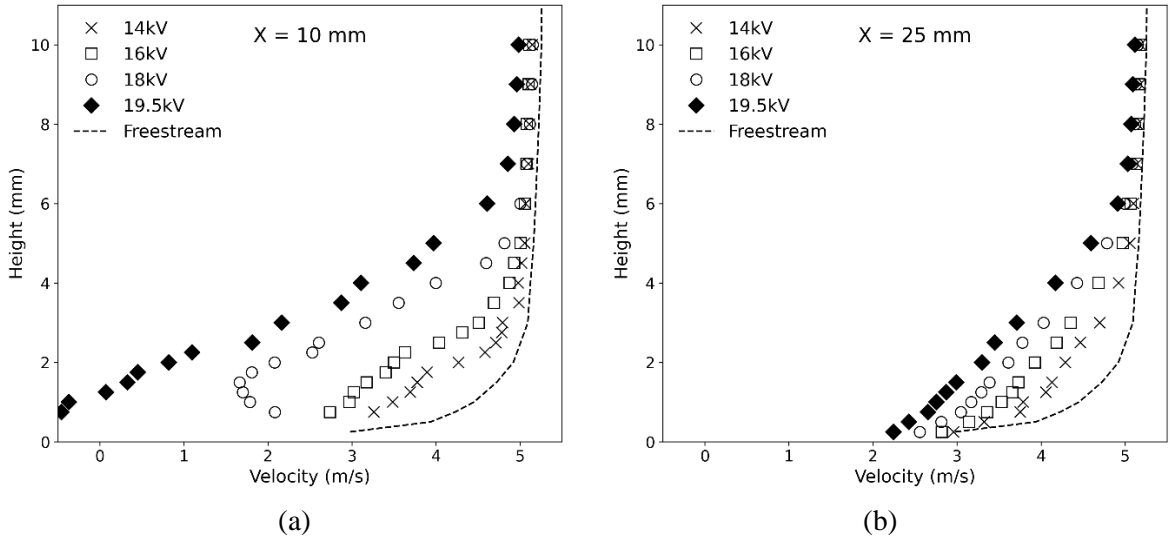

(a)                                                    (b)

**Figure 9. DBD actuator in $U_\infty$ = 5m/s counter-flow at x = 10 mm (a) and x = 25 mm (b) downstream. The DBD voltage is varied in the 14kV-19.5kV range, the AC frequency is set constant at 2kHz.**

For higher external flow, the effects of the DBD jet are less dramatic. Within the boundary layer, the largest decrease in velocity in counter-flow with $U_\infty$ = 11 m/s is approximately 2.0 m/s at y = 0.5 mm and x = 10 mm. No separation was observed in the $U_\infty$ = 11 m/s counter-flow cases, though the plasma region was not probed. While the effects of the DBD in counter-flow at $U_\infty$ = 11 m/s are less dominant than for slower freestream experiments, the effects of the EHD wall jet are still more significant than in co-flow at the same $U_\infty$.

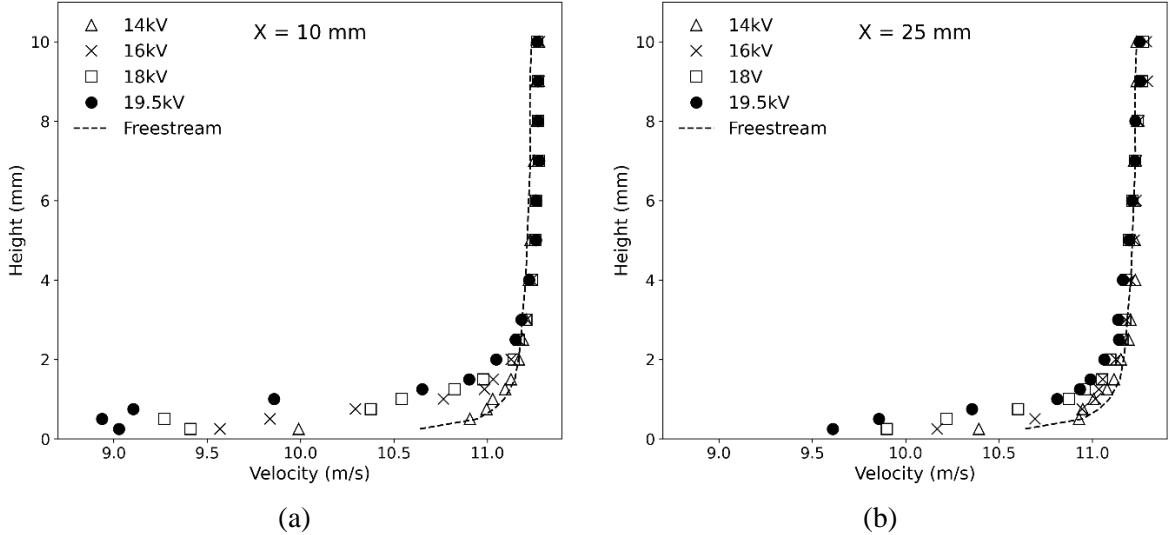

**Figure 10. DBD actuator in $U_\infty$ = 11 m/s counter-flow at x = 10 mm (a) and x = 25 mm (b) downstream. The DBD voltage is varied in the 14kV-19.5kV range, the AC frequency is set constant at 2kHz.**

The x-scans were performed while holding the y position constant to determine the boundaries of the separation region. Figure 11 shows the velocity profiles for y = 0.5 mm and y = 1.0 mm, while the x position was varied from x = 0 mm (edge of the active electrode) to x = 15 mm. The DBD voltage was $V_{AC}$ = 12, 14, 16, 18 kV at f = 2 kHz. The data for $V_{AC}$ = 19.5 kV is not shown due to the limited range of the pressure gauge.

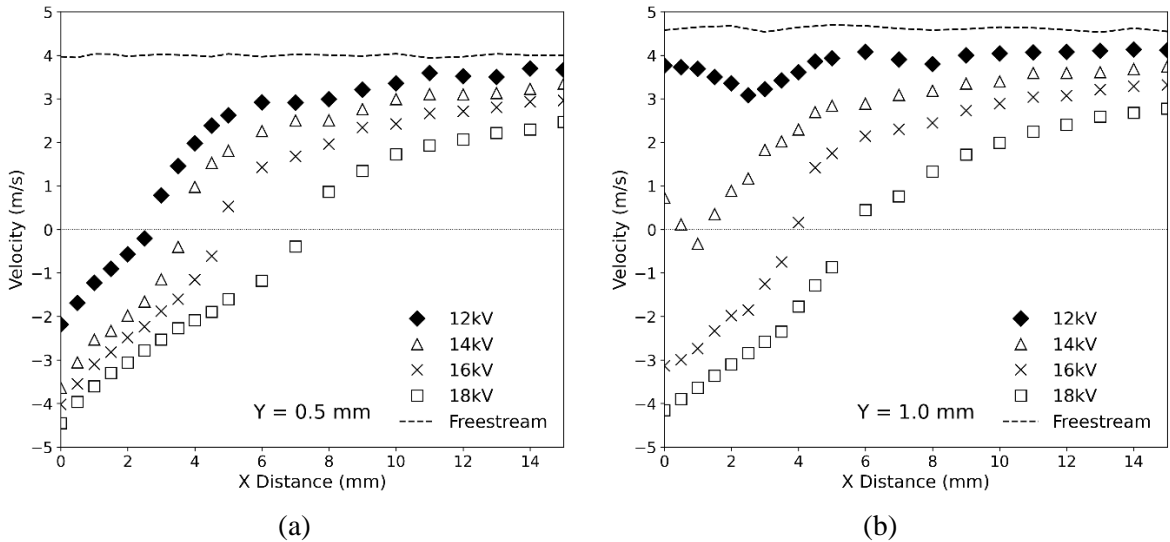

**Figure 11. DBD actuator in $U_\infty$ = 5 m/s counter-flow at y = 0.5 mm (a) and y = 1.0 mm (b). The DBD voltage is varied in the 14 kV-19.5 kV range, the AC frequency is set constant at 2 kHz.**

At y = 0.5 mm, immediately above the dielectric layer, separation is observed at all voltages. The x-location, where the velocity direction changes from backward to forward, determines the separation bubble's edge. At VAC = 18 kV, the separation length is approximately 7.5 mm downstream. At y = 1.0 mm, there is no signature of the separation bubble for VAC = 12 kV; however, it exists for the higher voltages.

To better visualize the flow pattern at 5 m/s, additional x-scans were performed, and the 2D velocity fields were reconstructed. Figure 12 shows the velocity contour plots for the counter-flow EHD jet obtained by merging x and y scans at $U_\infty = 5$ m/s. Each grid point in the figure is associated with a velocity measurement; the spatial resolution was 0.5 mm in both x and y directions, totaling approximately 400 measurements for each condition.

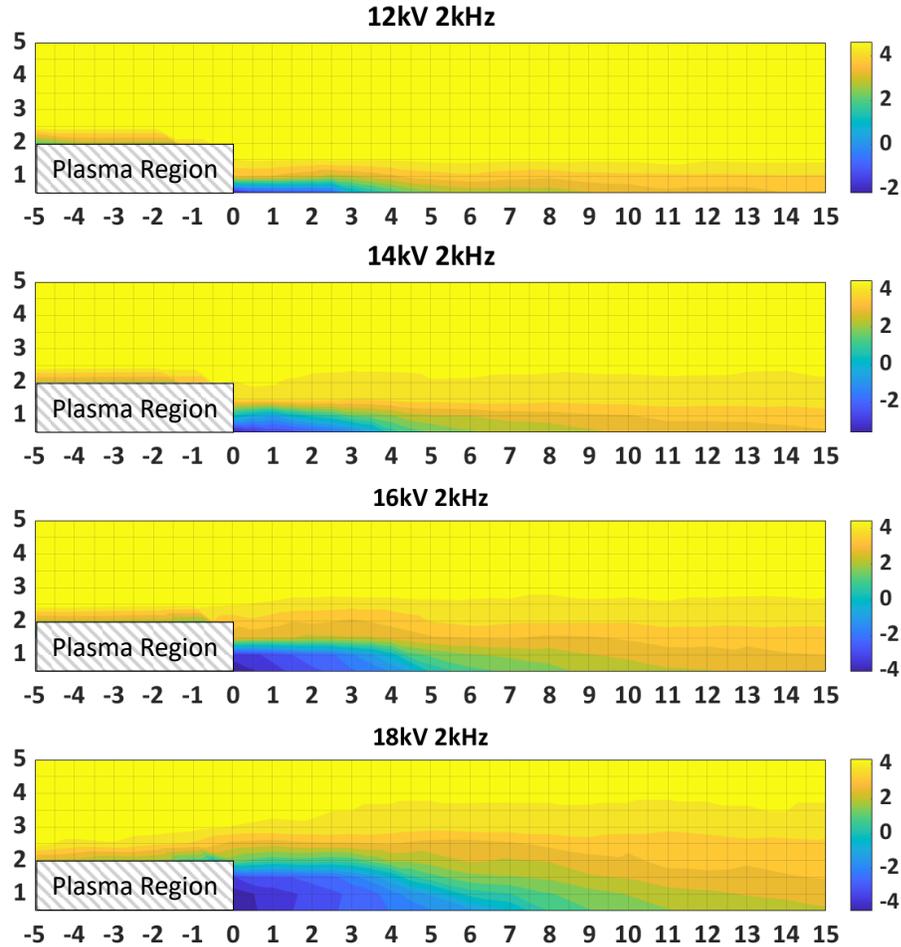

**Figure 12. X-velocity contour plot for EHD jet in counter-flow for $U_\infty = 5$ m/s for varying voltages. Gridlines correspond to recorded points spaced 0.5 mm apart in the x- and y-direction.**

All DBD actuations cause a separated region (for $U_\infty = 5$ m/s, RE=35,000) with a negative x-velocity downstream of the plasma injection. With the increase in DBD voltage, the edge of the reversed flow region within the separation bubble extends from 3.0 mm (12 kV) to >10.0 mm (18 kV) in the x-direction from the edge of the active electrode and from 0.6 mm (12kV) to > 1.75 mm (18 kV) in the y-direction. This growth in the length and height of the reversed flow region of the separated bubble is nonlinear with increasing voltage. The size and shape of the recirculation bubble are determined by the competition of the EHD jet strength vs. the forward momentum in the boundary layer. As the EHD injects negative momentum at high DBD voltages, it can overtake the momentum in the boundary layer at greater heights. Without sampling within the plasma region, it is challenging to characterize the entire length of the separation regions. It can be expected that the separation region extends into the forcing plasma region. Multiphysics CFD simulations can potentially address this issue; however, robust models need to be developed and validated. Although multiphysics CFD analysis is beyond the scope of this paper, preliminary results with a simplified momentum model proposed in [95] also suggest that the momentum injection in the counter-flow scenario triggers flow separation, though additional validation is required.

### 3.5. Momentum Difference

This section discusses the momentum difference in the boundary layer due to the EHD momentum injection. Momentum difference is calculated at x = 10 mm downstream of the DBD wall jet by integrating the velocity profiles in the y-direction up to a height where mass and momentum are injected and not entrained. Note that the x = 10 mm location is in the direction of the external flow with the datum at the plasma generation edge of the DBD actuator. Thus, in the co-flow case, the x = 10 mm location is downstream or in front of the plasma region and above the dielectric and grounded electrode (Figure 5); however, in the counter-flow case, the x = 10 mm location is behind the plasma region and above the active electrode (Figure 8).

Figure 4 compares the DBD with external flow cases against an EHD jet in a quiescent environment. The absolute value of the momentum difference is shown, as the momentum difference is calculated by subtracting the counter-flow actuation profile from the un-actuated boundary layer profile. The literature does not contain any momentum comparison between the co- and counter-flow DBD injections.

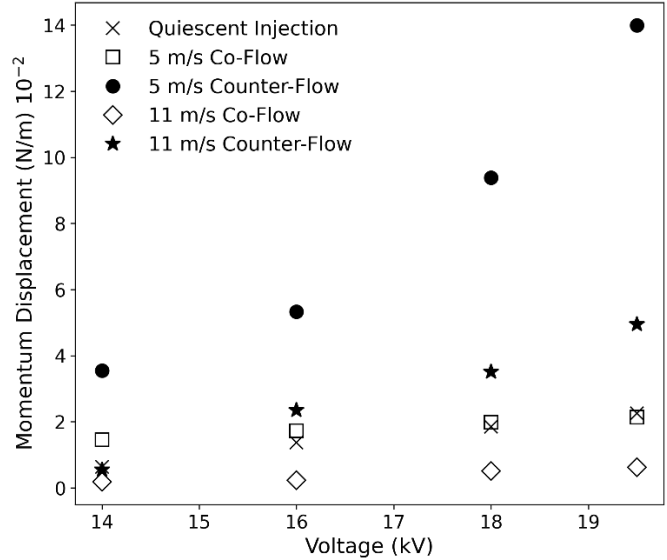

**Figure 13. DBD momentum difference at $U_\infty$ = 5 m/s and $U_\infty$ = 11 m/s external flow**

In the co-flow scenarios, the momentum addition into the boundary layer is equal to or lower than the momentum injected in the EHD jet. The momentum difference appears linear with $V_{AC}$; however, the change in total momentum is relatively flat, suggesting that momentum dissipation is driven primarily by the inner layer wall jet interaction with the wall. At 11 m/s co-flow, lower momentum differences are found for all voltages, and an increase in turbulent dissipation can explain this. The increase in dissipation is shown in the velocity profiles in Figure 6 and Figure 7 as the effect of the DBD momentum can still be seen at x = 25 mm downstream when $U_\infty$ = 5 m/s (Re = 35,000), but the effect of the DBD momentum is almost unseen at x = 25mm downstream when $U_\infty$ = 11 m/s (Re = 75,000). Unlike the experiments in quiescent conditions [95], the fluid momentum of the EHD jet in the co-flow injection is not conserved as it travels downstream. In the counter-flow configuration, the momentum difference is more significant due to reversing flow near the wall. The largest difference is observed in counter-flow, at $U_\infty = 5$ m/s. The momentum difference is ~ 6.5x greater than its co-flow counterpart.

The ratio of momenta within the DBD jet heights, $M^*$, is proposed as a nondimensional relation that could predict separation in the counter-flow injection. $M^*$ represents the DBD momentum injection compared to the inertial force in the external flow. $M^*$ is defined as:

$$M^* = \frac{M_{DBD}}{M_{External}} = \frac{\int_0^{h_{jet}} U_{quiescent\ DBD}^2(y) dy}{\int_0^{h_{jet}} U_{external\ flow}^2(y) dy} \quad (7)$$

The $M_{DBD}$ value and the height of the jet ($h_{jet}$) in a quiescent environment can be directly measured or estimated from the empirical relationship as proposed by Tang et al. [95]. E.g., the height of the DBD jet in Figure 4 varies from 2 mm to 2.25 mm at the range of the voltages tested. The $M_{External}$ value can be estimated analytically, numerically, or experimentally for a given value of the $h_{jet}$. The ratio of the terms can be evaluated to determine flow separation criteria $M^*$; the higher values are likely to result in flow separation. The values of $M^*$ in the 5 m/s and 11 m/s cases are shown in Table 1. In this limited set of experiments, the separation was observed for cases with $M^* > 0.1$ (for $M^* < 0.1$, counter-flow DBD did not induce separation). Additional testing and numerical simulations at various DBD and external flow conditions could further define separation threshold criteria. Note that MDBD varies with DBD parameters and the x-position within the jet as it expands and loses momentum due to viscous dissipation. At the same time, the value MExternal depends on the external flow conditions and the x-position of the DBD jet as the jet thickness changes with the x-position.

**Table 1. Conditions of Separation due to a DBD Jet in Counter-Flow**

| Reynolds | BL Height (mm) | DBD Momentum (mN/m) | M* | Separation |
|---|---|---|---|---|
| 35,000 | 8.0 | 6 – 22 | 0.14 – 0.52 | Yes |
| 75,000 | 2.5 | 6 – 22 | 0.02 – 0.07 | No |

### 4. MOMENTUM INJECTION MODEL

Numerical modeling of DBD has generally been categorized into three categories with increasing complexity: momentum injection models, simplified ion injection models, and species transport models. Momentum injection models such as that of Yoon *et al.* [34] and Kriegseis *et al.* [96] have been shown to accurately predict steady-state fluid flowfield of DBD actuators in a few configurations using empirically estimated forcing fields while remaining extremely computationally inexpensive. Simplified ion injection models such as the Orlov [113], Shyy [23], and the Suzen and Huang model [29] employ analytically or semi-empirically estimated charge density boundary conditions or charged density regions to model the transport of electrons and generalized ions. The electric potential and the charge density are then used to calculate the electrostatic Lorentz force with no magnetic field, and this EHD force is then coupled to the Navier-Stokes. In species transport models such as that of Bie *et al*. [114] and Soloviev *et al*. [115], the transport of the dominant chemical species, the resulting ions, and the radicals is modeled, and the distribution of ions is used to compute the electrostatic force at each time, similar to the simple ion models. Simplified ion and momentum injection models have been popular as they can readily be used for different applications while remaining relatively computationally lean.

An early implementation of a momentum injection model based on previously published empirical measurements is tested in co-flow and counter-flow. No published DBD model has been tested in an external co-flow or counter-flow. This supplemental information further supports the previously proposed momentum injection model in an external flow while providing insights into the fluid interactions at the focus of this manuscript. The two-dimensional schematic is identical to Figure 5 and Figure 8. The domain height is set to match the experimental wind tunnel of 10cm. The velocity profile of the wind tunnel is defined as a custom user-defined velocity profile. A mesh of 332,000 cells is employed with refinement near the forcing region, and courser meshes were tested to ensure mesh independence. Since the DBD actuator does not have a fluid mass flux, the resulting fluid governing equations are the incompressible Navier-Stokes continuity and momentum equation with an added momentum source, defined as

$$\nabla \cdot \mathbf{u} = 0 \qquad (8)$$

$$\rho \frac{D\mathbf{u}}{Dt} = -\nabla P + \mu \nabla^2 \mathbf{u} + \vec{f}_{EHD} \qquad (9)$$

The area of the momentum injection is within a right triangle region, similar to the approach of Shyy et al. [23], which assumes a linear approximation between plasma length and height. The author's previous work outlines that plasma length approaches ~ 8 mm for these electrical parameters, and the length-to-height ratio is approximately constant at L/H = 4. For all simulations, the steady-state k-ω turbulence model is used. A steady-state assumption for the DBD forcing is often assumed as the electrostatic and unsteady forcing timescales, and variances are considered sufficiently small. High-temporal resolution PIV data has shown that the time fluctuations of the DBD jet are often approximately 10% within a single voltage period; however, this variance depends on the applied voltage waveform [82, 116]. Yoon *et al.* used the one-equation Spalart-Allmaras turbulence model to model the DBD jet in quiescent flow; however, this significantly underpredicted the separation region in the counter-flow.

The resulting velocity profiles and momentum difference calculations are presented below in Figure 14. In the co-flow configuration, the momentum injection model matches the maximum velocity within 10%. However, the location of the maximum velocity in the model is higher than experimentally measured, and the velocity difference appears more compact. This is believed to be due to the inability of a two-dimensional CFD simulation to capture turbulence effects well. In addition, the higher location of the maximum velocity supports that the forcing distribution shape should likely be more concentrated at the near-wall region, possibly similar to the modified Gaussian used in Yoon et al. [34]. The total momentum difference matches the experimental measurement using the velocity profile integrated at x = 10mm.

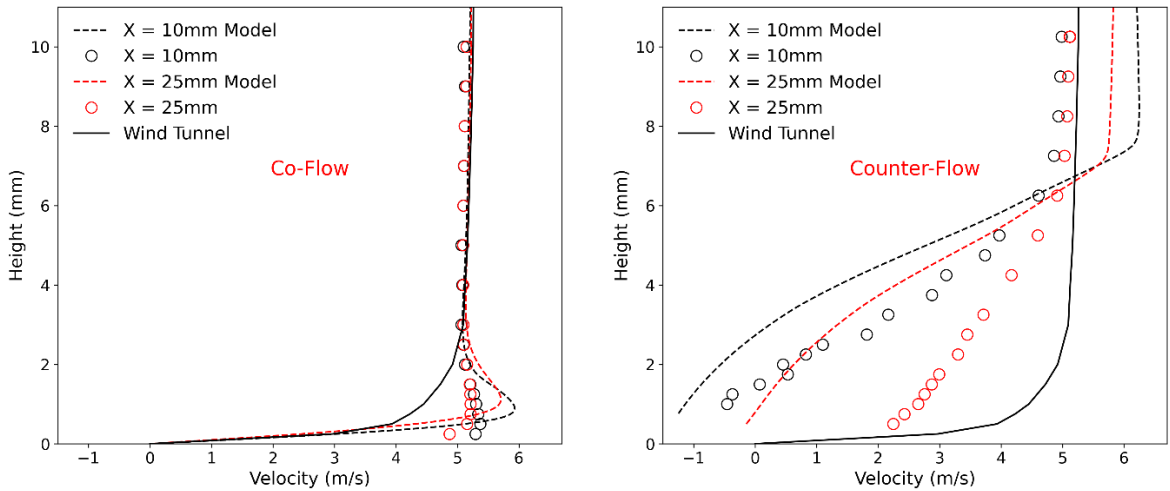

**Figure 14. Numerical and experimental DBD velocity profiles with 19.5kv 2kHz forcing (~22 mn/m) in 5 m/s external co- and counter-flow.**

In the counter-flow configurations, the strength of the separation region is overpredicted. This is believed to be due to the k-ω turbulence model incorrectly predicting the separation region, which is generally in line with many other adverse-pressure gradient studies, including one on the backward-facing step that over-predicted shear stress and reattachment locations [117]. The momentum deficit measured at the x = 10mm has an error of about 30%. In addition, experimental results show mass entrainment at higher locations in the wind tunnel compared to the numerical results that show entrainment as low as ~7mm above the plate. The downstream dissipation is not matched well, as the x = 25mm experimental profile shows higher dissipation than the numerical solution.

Overall, the presence of a separation region in the counter-flow case was the most challenging case to match. A basic numerical simulation will unlikely accurately predict flow in a boundary layer with a strong adverse pressure gradient inducing separation. Thus, advanced numerical efforts are needed to predict this DBD-induced separation region accurately. The main strength, yet main limitation, of this model, is its simplicity. However, that potential can only be fulfilled by further understanding the dependencies on turbulent models, unsteady forcing, and plasma forcing volume. More advanced turbulent models such as Large Eddy Simulation (LES), Detached Eddy Simulation (DES), or Direct Numerical Simulation (DNS) may be needed to resolve viscous effects, which is especially important in counter-flow correctly. Time-averaged forcing with a triangular plasma force shape may be appropriate for simple cases such as in co-flow, but in the cases such as the counter-flow or crossflow, shear stress and turbulent effects over the momentum deficit volume are magnified, and proper time-resolution may be required. With these aspects tackled, this model can serve as a valuable DBD design tool while providing accurate results and shedding important insight into the DBD forcing.

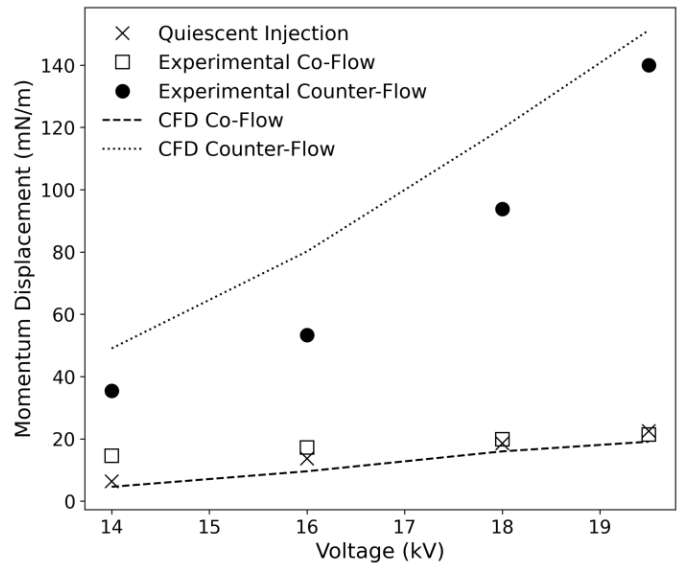

**Figure 15. Numerical and experimental DBD momentum difference with 19.5 kV 2 kHz forcing (~22 mN/m) in 5 m/s external co-flow and counter-flow**

# 5. CONCLUSION

We have experimentally investigated the performance of a DBD plasma actuator over a range of voltages (12 kV – 19.5 kV) at 2 kHz in co-flow and counter-flow with freestream velocities of 5 m/s and 11 m/s. The power consumption associated with DBD discharge is measured through capacitive measurements, with high temporal resolution throughout several cycles. For all voltages and freestream conditions in this experiment, there was no significant difference in power expenditure between the co-flow, counter-flow, and quiescent conditions, consistent with previous results. The DBD jet increased boundary layer velocity by > 2.0 m/s in co-flow and decreased the boundary layer velocity > 5 m/s in counter-flow (leading to fully reversed flow near the wall). The momentum difference in counter-flow leads to flow separation; separation zone boundaries and velocity magnitudes were evaluated using velocity magnitude contour plots.

At low freestream velocities, the EHD jet significantly influences boundary layer flow, and the dissipation is driven by the interaction of the DBD wall jet inner layer and the wall. However, at the higher freestream velocity, the external flow affects the outer layer of the EHD due to the more effective turbulent mixing. The counter-flow momentum difference is 6.5 times higher than its co-flow counterpart at $U_\infty = 5$ m/s. The momentum difference in counter-flow offers promising results for active flow control applications. A non-dimensional flow separation criteria $M^*$ is proposed as the ratio of DBD jet momentum to integrated boundary layer momentum. This experimental data set can be used to develop models and validate multiphysics simulations for EHD flow. Future research should extend the understanding of the relationship between the unsteady forcing of the DBD and the turbulent characteristics of the external flow.

# 6. ACKNOWLEDGMENTS

This work was funded by the Joint Center for Aerospace Technology Innovation (JCATI).

## NOMENCLATURE

| | |
|---|---|
| $C$ | Pitot tube correction factor |
| $E$ | Electric field |
| $f_{AC}$ | Frequency of the applied voltage |
| $\vec{f}_{EHD}$ | Electro-hydrodynamic force term |
| $i(t)$ | Current |
| $I_{dis}$ | Discharge current |
| $L$ | Spanwise Length |
| $M$ | The momentum of the induced jet |
| $P$ | Pressure reading from the pitot tube |
| $W$ | Discharge energy consumption |
| $W_{mech}$ | Mechanical power |
| $W_{elec}$ | Electrical power |
| $U(y)$ | Velocity at y height |
| $U_{max}$ | Maximum velocity of the wall jet |
| $U_\infty$ | External flow velocity |
| $V_{AC}$ | AC Voltage in the DBD actuator |
| $v$ | Time-averaged velocity |
| $t^*$ | Normalized time value |
| $\rho$ | Density |
| $Q$ | Mass flow rate |